\documentclass[letterpaper,12pt]{article}
\usepackage{amsmath}
\usepackage{amsfonts}
\usepackage{amssymb}
\usepackage{graphicx}
\usepackage{setspace}
\usepackage{authblk}
\usepackage{subfigure}

\author[1]{Faustino S\'anchez-Gardu\~no} 
\author[1,2]{Pedro Miramontes}
\author[3]{Tatiana T. M\'arquez-Lago \thanks{Corresponding author: tatiana.marquez@oist.jp}}

\affil[1]{Facultad de Ciencias, Universidad Nacional Aut\'onoma de M\'exico, Circuito Escolar. Cd. Universitaria, 04510, M\'exico D.F., M\'exico}
\affil[2]{Centro de Ciencias de la Complejidad, Universidad Nacional Aut\'onoma de M\'exico, Circuito Escolar. Cd. Universitaria, 04510, M\'exico D.F., M\'exico}
\affil[3]{Integrative Systems Biology Unit, Okinawa Institute of Science and Technology, Kunigami, Okinawa 904-0412, Japan}

\title{Role reversal in a predator-prey interaction}

\begin{document}

\maketitle

\begin{abstract}
Predator-prey relationships are one of the most studied interactions in population ecology. However, little attention has been paid to the possibility of role exchange between species once determined as predators and preys, despite firm field evidence of such phenomena in the nature. In this paper, we build a model capable of reproducing the main phenomenological features of one reported predator-prey role-reversal system, and present results for both the homogeneous and the space explicit cases. We find that, depending on the choice of parameters, our role-reversal dynamical system exhibits excitable-like behaviour, generating waves of species' concentrations that propagate through space.  

\end{abstract}


\section{Introduction}
\label{intro}
In 1988, A. Barkai and C.D McQuaid reported a novel observation in population ecology while studying benthic fauna in South African shores \cite{barkai}: a predator-prey role reversal between a decapod crustacean and a marine snail.  Specifically, in Malgas Island, the rock lobster {\it Jasus lalandii} preys on a type of whelk, {\it Burnupena papyracea}. As could be easily expected, the population density of whelks soared upon extinction of the lobsters in a nearby island (Marcus island, just four kilometers away from Malgas). However, in a series of very interesting controlled ecological experiments, Barkai and McQuaid reintroduced a number of {\it Jasus lalandii} in Marcus Island, to investigate whether the equilibrium observed in the neighboring Malgas Island could be restored. The results were simply astounding: 

\begin{quotation}
``The result was immediate. The apparently healthy rock lobsters were quickly overwhelmed by large number of whelks. Several hundreds were observed being attacked immediately after release and a week later no live rock lobsters could be found at Marcus Island.''
\end{quotation}

Surprisingly, and despite observations such as the report in \cite{barkai}, Theoretical Population Biology has largely ignored the possibility of predators and preys switching their roles. Of importance, the paper of Barkai and McQuaid suggests the existence of a threshold control parameter responsible for switching the dynamics between (a) a classical predator-prey system with sustained or decaying oscillations, and (b) a predator (the former prey) driving its present-day prey to local extinction.

It is worth noting there are some papers in the literature describing ratio-dependent predation (see, for example \cite{seitz} and \cite{holbrook}), but they are not related to the possibility of role-reversals. On the other hand, the likelihood of changing ecological roles as a result of density dependence has already been documented for the case of mutualism by Breton \cite{breton} and, in 1998, Hern\'andez made an interesting effort to build a mathematical scheme capable of taking into account the possible switches among different possible ecological interactions \cite{hernandez}. So, to the best of our knowledge, there are no theoretical studies --supported by field evidence-- specifically addressing predator-prey role-reversals yet.

\section{Mathematical model}

Predator-prey systems are generally modeled by adopting one of the many variations of the classical Lotka-Volterra model:

\begin{align}
\label{E0}
\dot{x} & = \alpha x-\beta xy\nonumber\\
\dot{y} & =-\gamma y+\delta xy,
\end{align}

\noindent where $\alpha$ denotes the intrinsic preys' rate of growth, $\beta$ corresponds to the rate of predation upon preys, $\gamma$ stands for the predators' death rate in absence of preys, and $\delta$ represents the benefit of predators due to the encounters with preys. Our goal is to assess whether modeling the role-reversal behavior observed by Barkai \& McQuaid \cite{barkai} is possible, when adopting appropriate parameters and assumptions. 

For instance, if one considers quadratic density dependence in the preys as well as in the predators, non-constant rates of consumption of preys by the predators, and the profiting of predators by the existence of preys, then it is possible to  suggest the following system:

\begin{align}
\label{E1}
\begin{array}{ccc}
\dot{x} & = & Bx(A/B-x)-C(x)xy\\
\dot{y} & = & -Dy-Ey^2+F(x)xy,
\end{array}
\end{align}

\noindent where $B$ represents the intrinsic growth rate of the prey in the absence of predators, $A/B$ the carrying capacity of the prey's habitat, $C(x)$ the rate of preys consumption by the population of predators, $D$ the predators' decay rate in the absence of preys, $E$ the intraspecific rate of competition among predators and, finally, $F(x)$ the factor of predator's profiting from preys. The ratio $F(x)/C(x)$ is then the fraction of prey biomass that is actually converted into predator biomass. The latter should remain constant, since the fraction of preys' biomass converted to predators' biomass is a physiological parameter, rather than a magnitude depending on demographical variables.

Thus, a particular case of system (\ref{E1}) in the appropriate rescaled variables is:

\begin{equation}
\label{E2}
\begin{array}{ccc}
\dot{x} & = & bx(1-x)-cx(k-x)y\equiv f(x,y)\\
\dot{y} & = & -ey(1+y)+fx(k-x)y\equiv g(x,y),
\end{array}
\end{equation}

\noindent where all the parameters are positive and $0<k<1$. In fact, all of the parameters have a relevant ecological interpretation: $b$ is the normalized intrinsic growth rate of the species with density $x$, $c$ is a measure of the damage intensity of the second species on the first one, $e$ is the normalized rate of predators decay and $f$ is the benefit (damage) the second population gets from the first one. Note the crucial role played by the interaction term $x(k-x)$, where $k$ stands for the first population threshold to switch from being prey to predator.

\section{Phase portrait analysis}

\subsection{Nullclines and equilibria}

The horizontal nullcline of the system of equations (\ref{E2}), that is $[b(1-x)-c(k-x)y]x=0$, has two branches: the vertical axis and the nontrivial branch

\begin{equation}
\label{E3}
y_h(x)=\frac{b(1-x)}{c(k-x)},
\end{equation}

\noindent which is a symmetric hyperbola with asymptotes: $x\equiv k$ and $y=b/c$ (See Figure 1). 

\begin{figure}
\label{fig:1}
 \includegraphics[width=120mm,height=95mm]{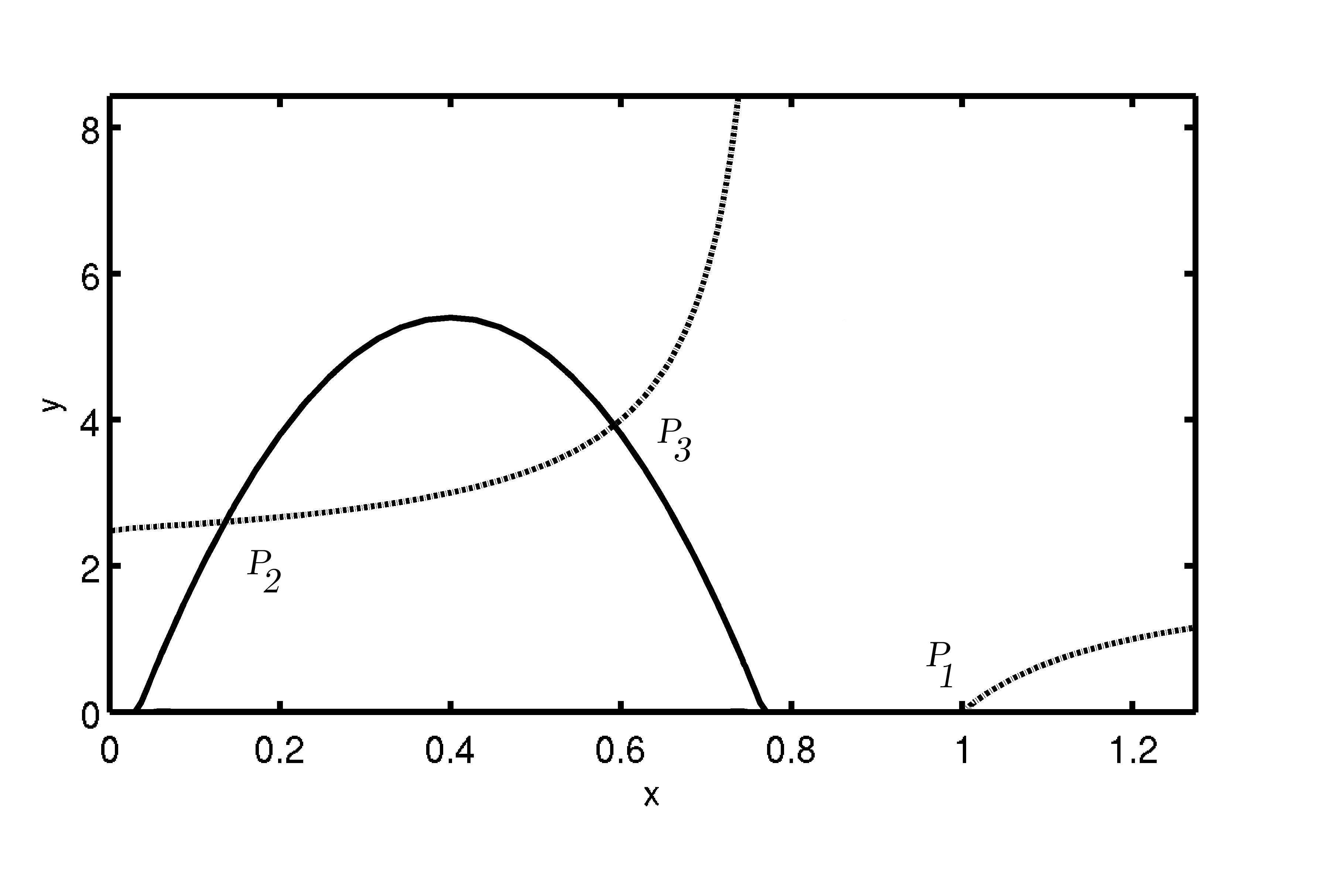}
 \caption{{\small Nullclines of the system of equations (\ref{E2}). The dotted line and the vertical axis are the horizontal nullcline. The continuous downward facing parabola and the horizontal axis are the vertical nullcline. $P_1$, $P_2$ and $P_3$ are the non-trivial equilibria. The origin of coordinates is a trivial equilibrium.}}
\end{figure}

\noindent The vertical nullcline, $[-e(1+y)+fx(k-x)]y=0$, also has two branches: the horizontal axis and

\begin{equation}
\label{E4}
y_v(x)=\frac{fx(k-x)}{e}-1,
\end{equation}

\noindent which is a parabola with $y_v(0)=y_v(k)=-1$, attaining its maximum at $x=k/2$, the value of which is

\[
y_v(k/2)=\frac{fk^2}{4e}-1.
\]

This term is positive if and only if $fk^2>4e$. The zeros, $x_1$ and $x_2$ of Equation (\ref{E4}) are given by

\[
x_1,x_2=\frac{f\pm\sqrt{f^2k^2-4fe}}{2f}.
\]

The latter are real numbers if and only if $fk^2\geq 4e$. The rate of change of $y_v$ is then

\[
y'_v(x)=\frac{f}{e}(k-2x).
\]

To analyze the system while keeping in mind the ecological interpretation of the variables and parameters, we will now consider the left branch of the horizontal nullcline (\ref{E3}), $fk^2>4e$ with $0<k<1$ and the region of the phase plane of the system of equations (\ref{E2}) defined as

\[
\mathcal R=\left\{(x,y)|0\leq x\leq 1,\;\;0\leq y<+\infty\right\}.
\]

The system of equations (\ref{E2}) has the equilibria: $P_0=(0,0)$, $P_{1}=(1,0)$ plus those states of the system stemming from the intersection of the nullclines $y_h$ and $y_v$ in the region $\mathcal R$. Such equilibria are defined by the $x$ in the interval $(0,1)$ satisfying the identity

\begin{equation}
\label{E5}
\frac{b(1-x)}{c(k-x)}=\frac{fx(k-x)}{e}-1,
\end{equation} 

or, equivalently, the $x$ that are roots of the third order polynomial

\begin{equation}
\label{E6}
F(x)=Ax^3+Bx^2+Cx+D,
\end{equation}

where $A=fc$, $B=-2fck$, $C=fck^2+be+ec$ and $D=-be-eck$.

The calculation of the nontrivial equilibria of (\ref{E2}) follows from the determination of the roots of (\ref{E6}). Consequently, due to the qualitative behavior of the functions $y_h$ and $y_v$ on $\mathcal R$, we are faced with the following possibilities:

\begin{enumerate}
\item The nontrivial branches of the nullclines do not intersect each other in the region of interest. In such a case, the system (\ref{E2}) has just two equilibria: $P_0$ and $P_1$ in ${\mathcal R}$. Figure 2a shows the relative position of the nullclines in this case, and Figure 3a the phase portrait of the system.

For fixed positive values of $b,c,e$ and $f$, and $k\in(0,1)$ such that $(fk^2/4e)>1$ one can see that both nullclines become closer with increasing values of $k$.

\begin{figure}[ht]
	\centering
	\begin{subfigure}
		\centering
		\includegraphics[width=2.4in]{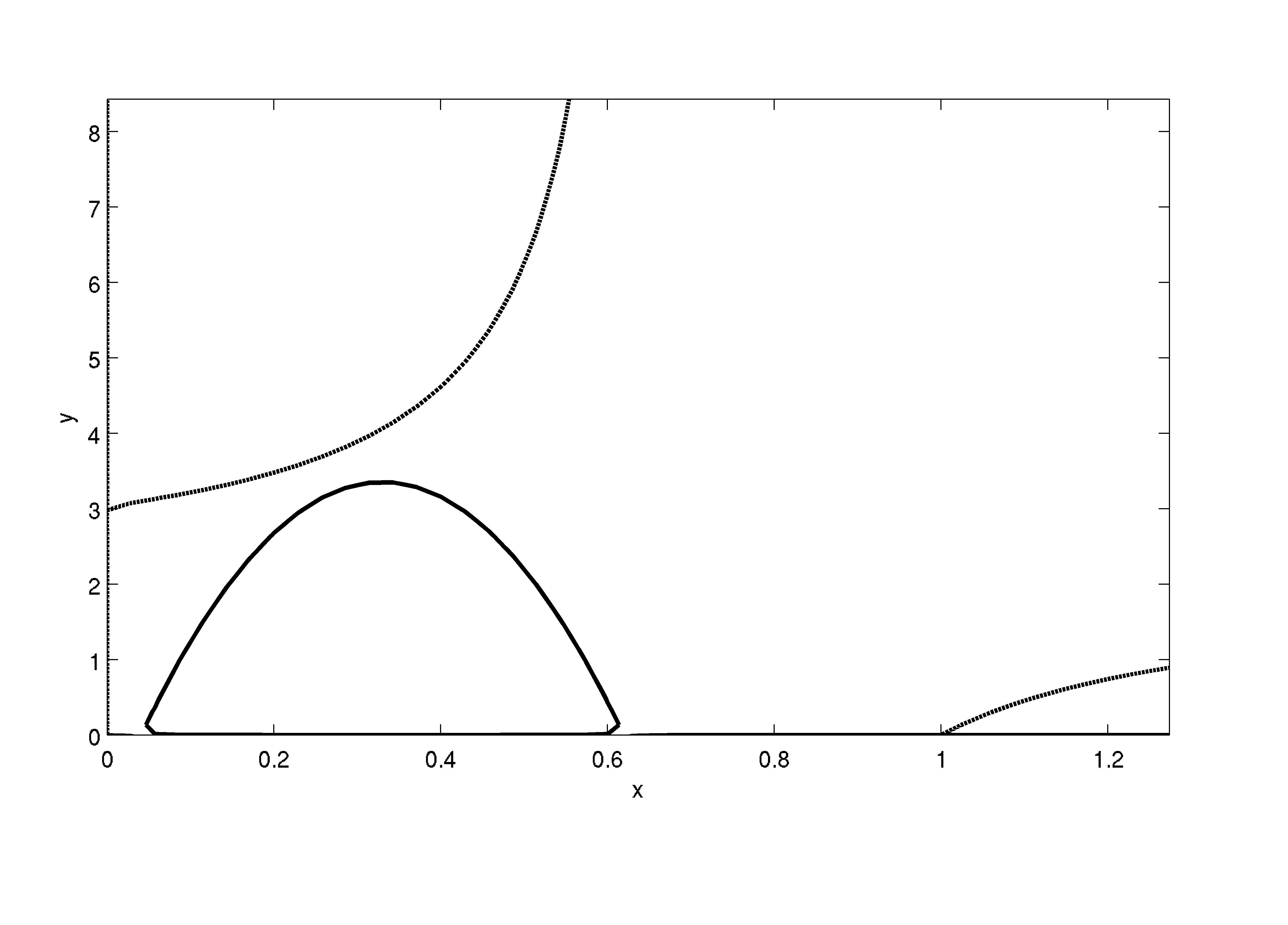}
		\label{fig:1a}		
	\end{subfigure}
	\quad
	\begin{subfigure}
		\centering
		\includegraphics[width=2.4in]{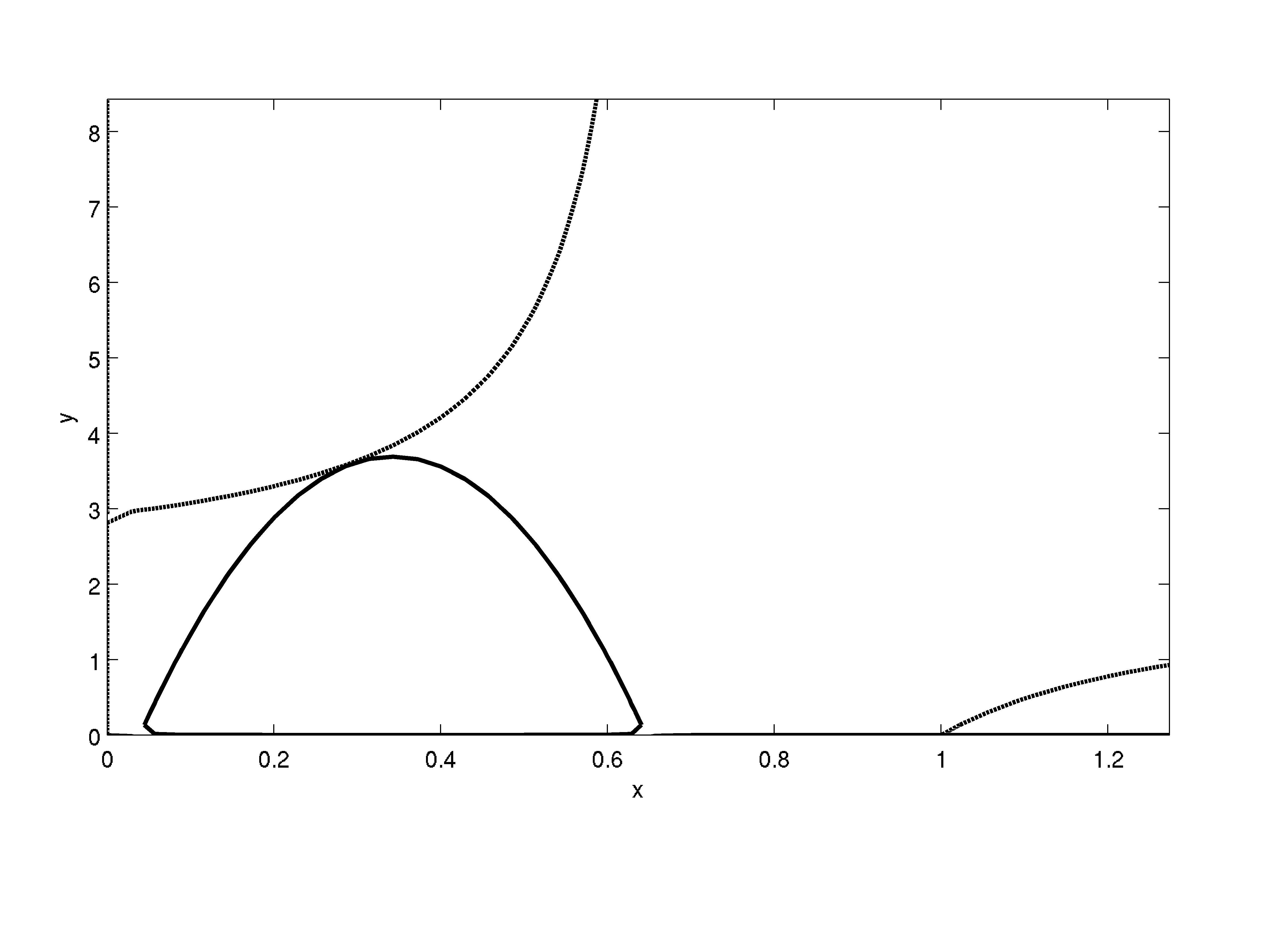}
		\label{fig:1b}
	\end{subfigure}
	\caption{The relative position of the nullclines can be parametrically controlled. In the left figure the downward facing parabola does not intersect the upper branch of the hyperbola. After a small change in the parameter $k$, both nullclines touch tangentially. Further changes in the parameter lead to a saddle-node bifurcation, and to the two transversal intersections depicted in Figure 1.}
\end{figure}

\begin{figure}[ht]
	\centering
	\begin{subfigure}
		\centering
		\includegraphics[width=2.4in]{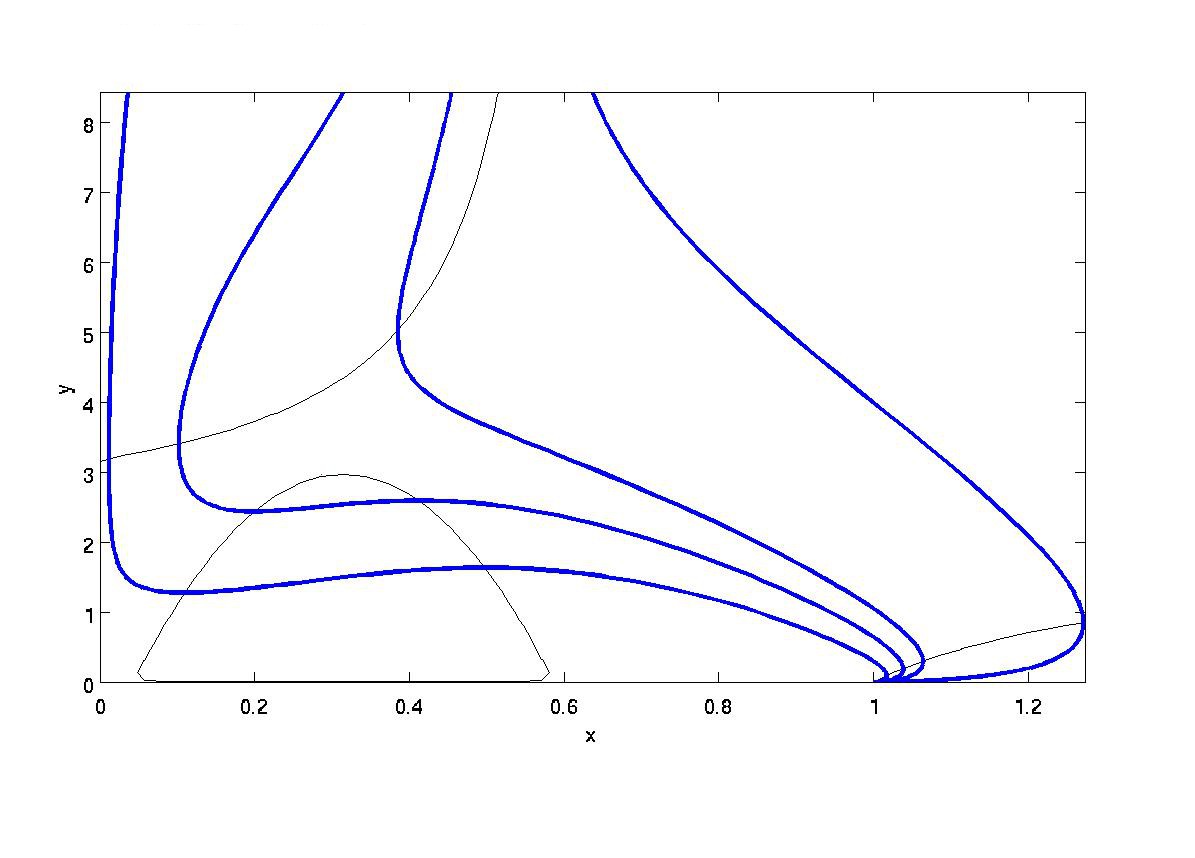}
	\end{subfigure}
	\quad
	\begin{subfigure}
		\centering
		\includegraphics[width=2.4in]{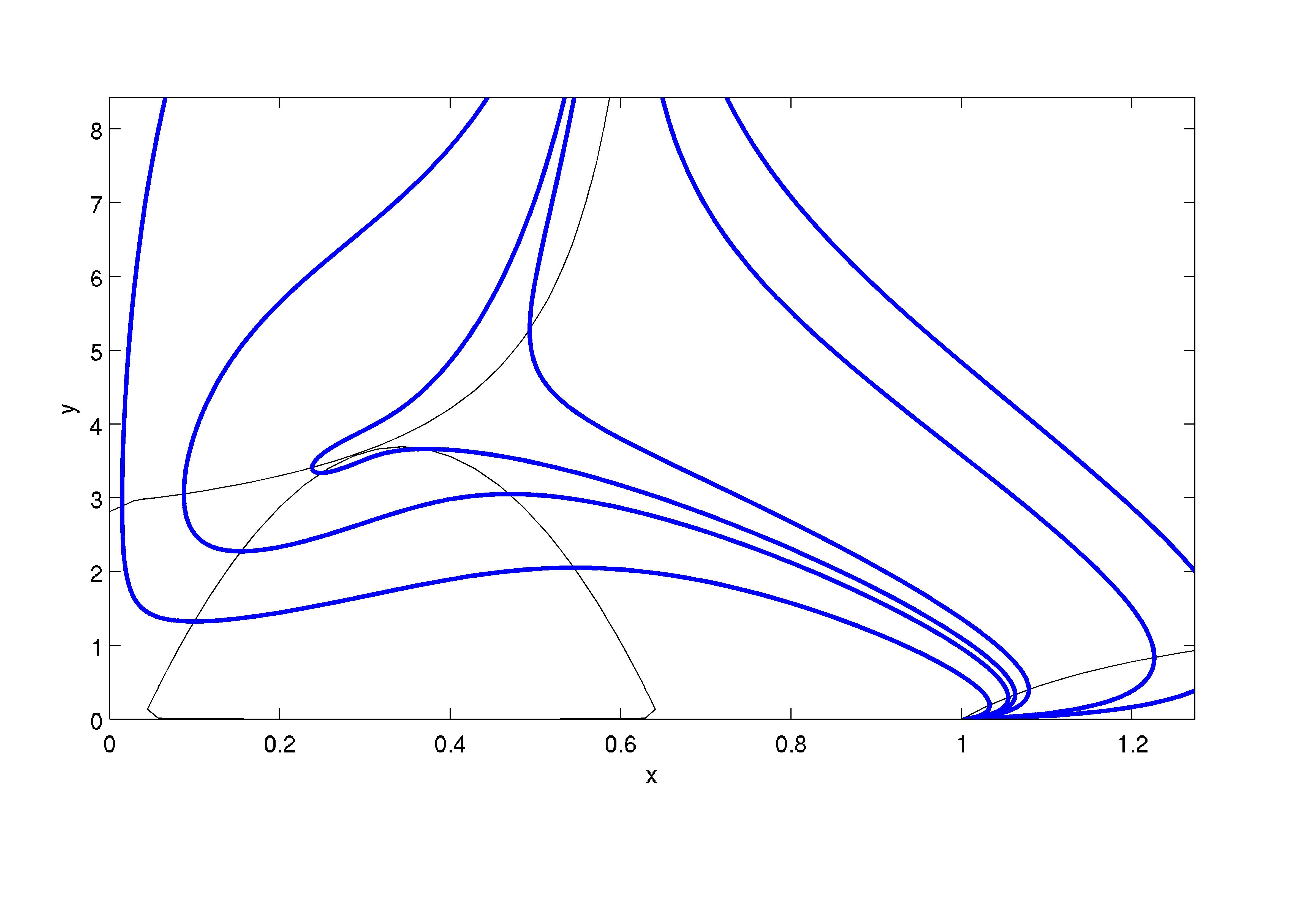}
	\end{subfigure}
	\caption{Phase portrait of the system. When the nullclines do not intersect, as seen in the left figure, the origin is an unstable node and there is only a nontrivial equilibrium on the x-axis which is stable. All the initial condictions lead to the lobsters extinction. However, when there is a tangential contact between the nullclines, as in the right figure, there is a new equilibrium, which is a degenerate node.}
\end{figure}

\item The nullclines $y_h$ and $y_v$ touch each other tangentially at the point $P^{*}=(x^*,y^*)$ in the region ${\mathcal R}$. Again, Figure 2b shows the relative position of the nullclines in this case, and Figure 3b the phase portrait of the system. In such a case $x^*$, in addition to satisfy (\ref{E5}), must also satisfy the condition $y_h'(x)=y_v'(x)$ {\em i.e.},

\begin{equation}
\label{E7}
\frac{b}{c}\frac{(1-k)}{(k-x)^2}=\frac{f}{e}(k-2x).
\end{equation}

If one assumes the existence of $x^*$ satisfying (\ref{E5}), the required extra condition (\ref{E7}) imposes the restriction $0<x^*<k/2$ on $x^*$, due to the positiveness of its left hand side. Moreover, from a geometrical interpretation of (\ref{E7}) it follows that: 

\begin{itemize}
\item (i) If

\[\frac{fk}{e}<\frac{b}{c}\frac{(1-k)}{k^2},\]
there is not any $x\geq 0$ such that $y'_h(x)=y'_v(x)$.  

\item (ii) If 
\[\frac{fk}{e}=\frac{b}{c}\frac{(1-k)}{k^2},\] 
the condition (\ref{E7}) is satisfied just at $x=0$.

\item (iii) If 
\[\frac{fk}{e}>\frac{b}{c}\frac{(1-k)}{k^2},\]
there exists exactly one value, $x^{*}\in(0,k/2)$, of $x>0$ such that the equality (\ref{E7}) holds. 

\end{itemize}

In any case, the point $P^*$ is a {\em non-hyperbolic} equilibrium of the system of equations (\ref{E2}). In fact, the proof that a tangential contact of the nullclines results in a point where the determinant of the Jacobian matrix of the system vanishes follows immediately, implying that at least one of its eigenvalues is zero.

\item The nullclines intersect each other transversally at two points, $P_2$ and $P_3$, belonging the region ${\mathcal R}$. For reference, please refer to Figure 1. In this case the system of equations (\ref{E2}) has two extra equilibria which arise from the {\em bifurcation} of $P^*$.

Here, if in addition to choosing the parameters $f$, $k$ and $e$ such that $fk^2/4e>1$, we select the rest of them such that:

\begin{itemize}

\item (i) $y_h(k/2)>y_v(k/2)$, i.e.,

\[\left[\frac{fk^2}{4e}-1\right]>\frac{b}{c}(2-k),\] 

guaranteeing the existence of the equilibria $P_2=(\tilde{x}_2,\tilde{y}_2)$ and $P_3= (\tilde{x}_3,\tilde{y}_3)$ above mentioned. Moreover, the coordinates of these points satisfy $0<\tilde{x}_2<k/2$, $k/2<\tilde{x}_3<k$, $\tilde{y}_i>0$ with $\tilde{y}_2<\tilde{y}_3$. Here $i=2,3$.

\item (ii) $y_h(k/2)=y_v(k/2)$, i.e.,
\[\left[\frac{fk^2}{4e}-1\right]=\frac{b}{c}(2-k).\]

Here we have $P_2=(\tilde{x}_2,\tilde{y}_2)$ with $0<\tilde{x}_2<k/2$ and $0<\tilde{y}_2<\frac{b}{c}(2-k)$. Meanwhile, $P_3= (k/2,\frac{b}{c}(2-k))$.
   
\end{itemize}
\end{enumerate}

\subsection{Local dynamics}

Part of the local analysis of the system of equations (\ref{E2}) is based on the linear approximation around its equilibria. Thus, we calculate the Jacobian matrix of the system (\ref{E2}):

\begin{equation}
\label{E9}
J[f_1,f_2]_{(x,y)} = \left[\begin{array}{cc}
b(1-2x)-cy(k-2x) & -cx(k-x)\\
fy(k-2x) & -e-2ey+fx(k-x)\end{array}
\right].
\end{equation}

By a straightforward calculation, we obtain the eigenvalues of the Jacobian matrix (\ref{E9}) at the point $P_0$. These are: $\lambda_1=b>0$ and $\lambda_2=-e<0$. Hence, $P_0$ is saddle point of the system (\ref{E2}), for all positive parameter values. By carrying out similar calculations we obtain the corresponding eigenvalues of matrix (\ref{E9}) at $P_1$, which are: $\lambda_1=-b<0$ and $\lambda_2=f(k-1)-e$. The restriction $0<k<1$ on $k$ implies that $(f(k-1)-e)<0$. Therefore, $P_1$ is an asymptotically stable node for all the positive parameter values appearing in system (\ref{E2}). 

Now we carry out the local analysis of (\ref{E2}). We notice two cases, depending on the relative position of the nullclines:

\vspace{4mm}
\noindent
{\bf Case 1.} The main branches (\ref{E3}) and (\ref{E4}) of the nullclines do not intersect on ${\mathcal R}$.

Here, any trajectory of system (\ref{E2}) starting at the initial condition $(x_0,y_0)$ with positive $x_0$ and $y_0$ tends to the equilibria $P_0$ as time goes to infinity. Thus, the region ${\mathcal R}^{+}$ is the basin of attraction of $P_1$. Invariably, the species with density $y$ vanishes, implying non coexistence among the interacting species. Meanwhile, the other species approach the associated carrying capacity.

\vspace{4mm}
\noindent
{\bf Case 2.} The nullclines intersect each other at the points $P_2=(\tilde{x}_2,\tilde{y}_2)$ and $P_3=(\tilde{x}_3,\tilde{y}_3)$, where none is tangential. Here $\tilde{x}_2$ and $\tilde{x}_3$ satisfy $0<\tilde{x}_2<k/2$ and $k/2<\tilde{x}_3<k$. In a neighborhood of $P_2$ and $P_3$, the functions $f_1$ and $f_2$ satisfy the Implicit Function Theorem. In particular, each one of the identities $f_1(x,y)=0$ and $f_2(x,y)=0$ define a function there. Actually, these are $y_h(x)$ and $y_v(x)$ given in (\ref{E3}) and (\ref{E4}), respectively. Their derivative at $\tilde{x}_i$ with $i=2,3$ is calculated as follows

\[y_h'(\tilde{x}_i)=-\frac{f_{1x}(\tilde{x}_i,\tilde{y}_i)}{f_{1y}(\tilde{x}_i,
\tilde{y}_i)}\;\;\mbox{and}\;\;y_v'(\tilde{x}_i)=-\frac{f_{2x}(\tilde{x}_i,
\tilde{y}_i)}{f_{2y}(\tilde{x}_i,\tilde{y}_i)}. \]

By using these equalities, we can state the following proposition.

\vspace{3mm}
\noindent
{\bf Proposition 1.} {\em The equilibrium $P_2$ is not a saddle point. Meanwhile, the equilibrium $P_3$ is a saddle point for all the parameter values.}

\vspace{3mm}
\noindent
A proof of this proposition and some remarks can be found in Appendix A.

\subsection{Global analysis}

As we have already shown, system (\ref{E2}) has four equilibrium points. These are illustrated in Figure 4. The origin is a saddle point with the horizontal and vertical axis as its unstable and stable manifolds. $P_1$ and $P_2$ are, respectively, a node and a saddle for parameter values after the bifurcation, and $P_3$ is a stable node. The stable manifold of the saddle point is a separatrix dividing the phase space in two disjoint regions: the set of initial conditions going to $P_1$, and the complement with points going to $P_3$. Moreover, our numerical solution shows the existence of an homoclinic trajectory starting and ending in the saddle point. Thus, we have a bistable system.

\begin{figure}[ht] 
\centering
 \includegraphics[width=120mm,height=95mm]{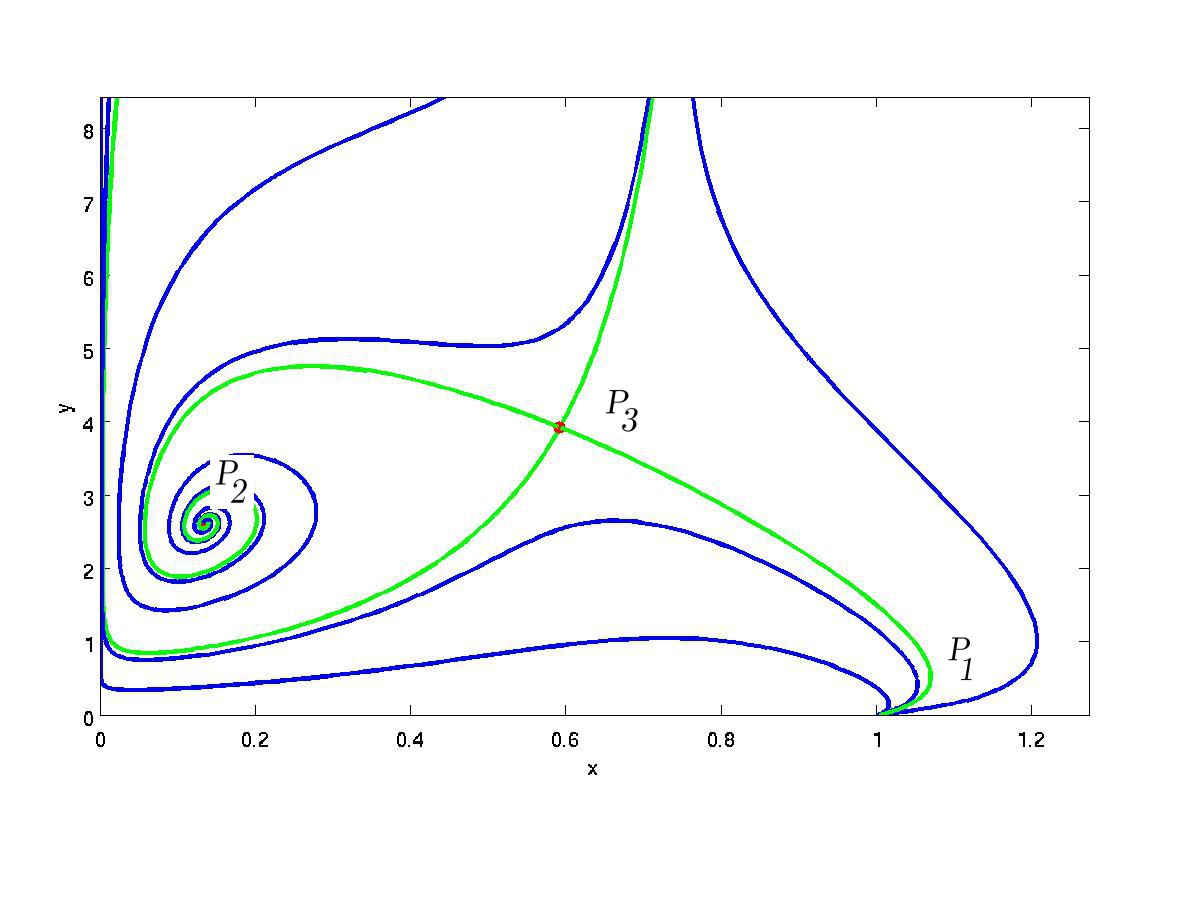}
 \caption{{\small The equilibria of system (\ref{E2}). From left to right: $P_2$ is a stable node, $P_3$ is a saddle and $P_1$ is another stable node. The heteroclinic trajectory joining the saddle point to the stable node is easily identified. The stable manifold is a separatrix between the basin of attraction of $P_1$ and $P_2$.}}
\end{figure}

The bistability of system (\ref{E2}) has an interesting ecological interpretation: the coexistence of the interacting species occurs whenever the initial population densities $(x_0,y_0)$ are located in the region above the saddle point unstable manifold. In this case, both populations evolve towards the attractor $P_2$. 

On the other hand, if the initial population densities $(x_0,y_0)$ are below the separatrix, the population densities $(x(t),y(t))$ evolve towards the equilibria $P_1$ implying the non-coexistence of the species and, invariably, the species with population density $y$ vanishes. The heteroclinic trajectory of system (\ref{E2}) connecting the saddle ($P_3$) with the node ($P_2$ -- or focus, depending on the set of parameters), in addition to the coexistence of the species, also tells us that this occurs by the transition from one equilibrium to another as time increases.

\section{Spatial dynamics}
  
To describe more accurately our role-reversal system, we extended our model of system (\ref{E2}) to incorporate the spatial variation of the population densities. Here, if we denote by $u(\vec{r},t)$ and $v(\vec{r},t)$ the population density of the whelks and lobsters at the point $\vec{r}$ at time $t$, the resulting model is:

\begin{align}
\label{E10}
\begin{array}{ccc}
u_t & = & D_u\nabla^2u+bu(1-u)-cu(k-u)v\\
v_t & = & D_v\nabla^2v-ev(1+v)+fu(k-u)v,
\end{array}
\end{align}


where the subscript in $u$ and $v$ denotes the partial derivative with respect the time, and $\nabla^2$ is the laplacian operator. Here, $D_u>0$, $D_v>0$ correspond to the diffusivity of the species with density $u$ and $v$, i.e. that of whelks and lobsters, respectively. It is worth noting that the original variables have been rescaled, but still denote population densities.

We then proceeded to construct numerical solutions of the system (10) in three different domains: a circle with radius 2.2 length units (LU), an annulus defined by concentric circles of radii 2.2 LU and 1 LU, and a square with side length of 4.6 LU. All domains were constructed to depict similar distances between Malgas island and Marcus island (roughly 4km). In the first one, the annular domain, we try to mimic the island habitat of whelks and lobsters as a concentric domain. The other two domains are used to confirm the pattern formation characteristic of excitable media, and to reject any biases from the shape of the boundaries. 

To obtain numerical solutions of all spatial cases, we used the finite element method with adaptive time-stepping, and assumed zero-flux boundary conditions. Accordingly, we discretized all spatial domains by means of Delaunay triangulations, until a maximal side length of 0.17 was obtained. The latter defines the approximation error of the numerical scheme. We attempted to describe two entirely different situations by using a single set of kinetic parameters: that of Malgas island, where both species co-exist, and Marcus island, where whelks soar and lobsters become extinct. The only difference between these two cases was the initial conditions used.

Aside, one could intuitively assume whelks motion to be very slow, or even negligible in comparison to that of lobsters. However, it is worth considering how slow, and whether fluid motion could aftect this speed. While there is no data specific to {\it Jasus lalandii} and {\it Burnupena papyracea} in islands of the Saldanha Bay, data of similar species can be found in the literature. For instance, a related rock-lobster species, {\it Jasus Edwardii} has been found to move at a rate of 5-7 km/day \cite{cobb}. In contrast, whelks within the superfamily {\it Buccinoidea} have been found to move towards food at rates between 50 and 220 meters/day (see \cite{himmelman} and \cite{lapointe}). Importantly, predation by whelks remains seemingly unaffected by variations in water flow \cite{powers}. 

By putting these findings together, we argue a reasonable model need not incorporate influences from shallow water currents, and would assume whelks to move toward `bait' at a speed roughly one order of magnitude smaller than that of lobsters. Thus, we opted for a two-dimensional habitat, and one order of magnitude difference between the non-dimensional isotropic diffusion rates ($D_u =  0.01$ and $D_v =  0.1$). Aside, our choice of reaction parameters was: $b =  10$, $c =  33.8$, $e =  0.5$, $f =  30$, and $k =  0.9$. Regarding initial conditions, we adopted the following scenarios, representing the different scenarios of weighted biomass:

\begin{enumerate}
\item Malgas Island: the initial density of whelks at each element was drawn from a uniform distribution 0.1 * U(0.25, 0.05), and that of lobsters from U(0.25, 0.05).
\item Marcus Island: the initial density of whelks at each element was drawn from a uniform distribution U(0.25, 0.05), and that of lobsters from 0.1 * U(0.25, 0.05).
\end{enumerate}

Results are shown in Figure 5, corresponding to averaged densities of whelks and lobsters in the three different spatial domains, respectively. Simulations in an annular domain can be found in the Supplementary Material.

\begin{figure}[h] 
\centering
 \includegraphics[width=120mm,height=95mm]{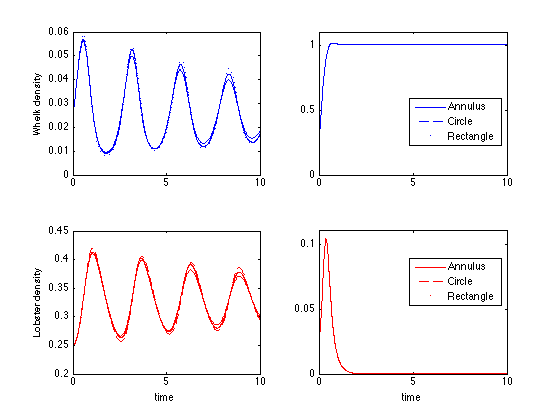}
 \caption{{\small Time evolution of whelk and lobster densities, in different spatial domains. Cases (a,c) correspond to initial conditions representing Malgas island, while (b,d) correspond to initial conditions representing Marcus island.}}
\end{figure}

Interestingly, changes in density are usually accompanied with wave-like spatial transitions in each species density. Examples of this spatial transient patterns can be found in Figures 6 and 7, for annular and rectangular domains in Malgas island and Marcus island, respectively.

\begin{figure}
\centering
 \includegraphics[scale=.5]{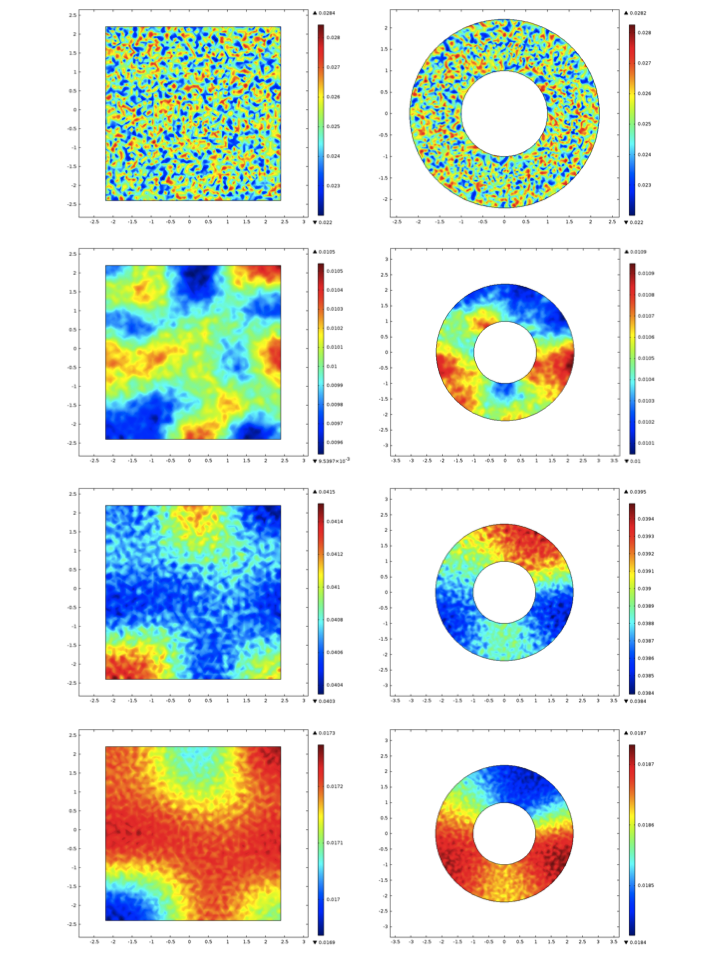} 
 \caption{{\small Time evolution of whelks' densities using initial conditions representing Malgas island, in annular (top) and rectangular (bottom) domains. Cases correspond to t = 0, 2, 6 and 10, from left to right.}}
\end{figure}

\begin{figure}
\centering
 \includegraphics[scale=.5]{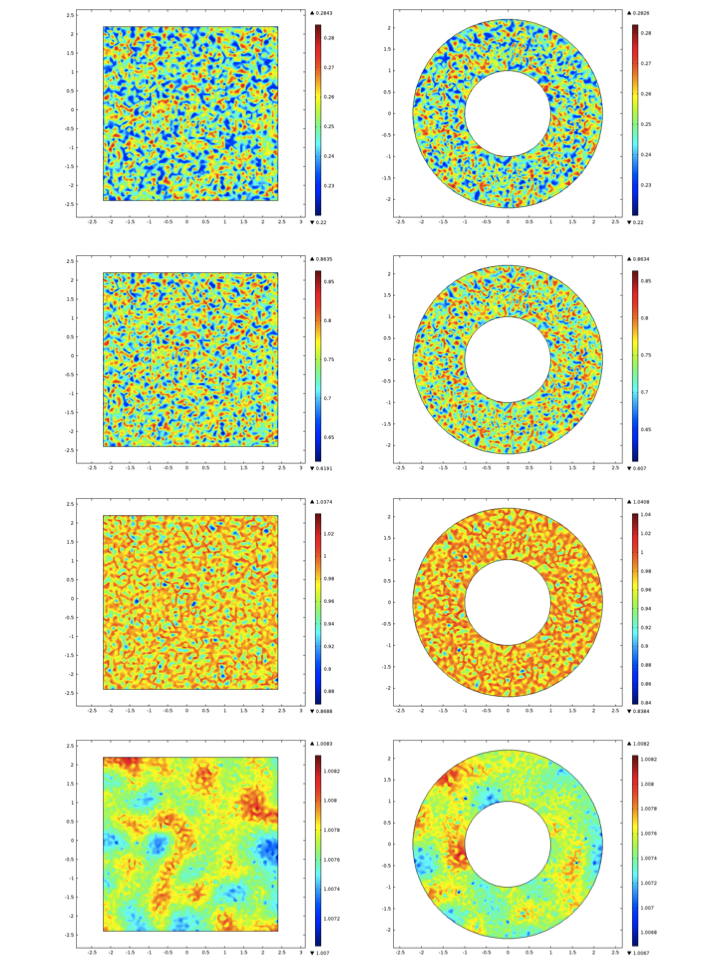} 
 \caption{{\small Time evolution of whelks' densities using initial conditions representing Marcus island, in annular (top) and rectangular (bottom) domains. Cases correspond to t = 0, 0.25, 0.5 and 1, from left to right.}}
\end{figure}

\section{Discussion and final remarks}

We have modeled a well documented case of role-reversal in a predator-prey interaction. Our model pretends to capture the essential ecological factors within the study of Barkai and McQuaid \cite{barkai}, who did an extraordinary field work and meticulously reported this striking role-reversal phenomenon happening between whelks and lobsters in the Saldanha Bay. 

The analysis of our model and corresponding numerical solutions clearly predict the coexistence of both populations and the switching of roles between the once denoted predators and preys. Here, the coexistence scenario corresponds to the case when lobsters predate upon whelks, and role-reversal corresponds to the case when whelks drive the population of lobsters to extinction, as observed by Barkai and McQuaid in the field. 

Moreover, by introducing spatial variables and letting both populations diffuse within a spatial domain, we obtain patterns that are characteristic of excitable media \cite{krinsky}. Of particular interest is the upper row of Figure 6, where self-sustained waves travel in the annular region. The latter is not entirely surprising, as the ordinary differential equation model in which the spatial case was based shows bistability. Nevertheless, our findings are novel in that, to the best of our knowledge, there are no reports of ecological interactions behaving as excitable media.

\section{Acknowledgements} PM was supported by UNAM-IN107414 funding, and wishes to thank OIST hospitality during last stages of this work. TML was supported by OIST funding. 

\newpage

\section{Appendix A}

{\bf Proof.} First we prove the second part of our Proposition. At $P_3$ we have

\[y_{h}'(\tilde{x}_{3})=-\frac{f_{1x}(\tilde{x}_3,\tilde{y}_3)}{f_{1y}(\tilde{x}_3,\tilde{y}_3)}>0,\]
then the partial derivatives $f_{1x}(\tilde{x}_3,\tilde{y}_3)$ and $f_{1y}(\tilde{x}_3,\tilde{y}_3)$ have opposite signs at that point but $f_{1y}=-c\tilde{x}_3(k-\tilde{x}_3)$ with $k/2<\tilde{x}_3<k$ resulting in $f_{1y}(\tilde{x}_3,\tilde{y}_3)<0$ hence $f_{1x}(\tilde{x}_3,\tilde{y}_3)>0$. Similarly 

\[y_{v}'(\tilde{x}_3)=-\frac{f_{2x}(\tilde{x}_3,\tilde{y}_3)}{f_{2y}(\tilde{x}_3,\tilde{y}_3)}<0,\]
implying that $f_{2x}$ and $f_{2y}$ have the same sign at $(\tilde{x}_3,\tilde{y}_3)$ but $f_{2x}(\tilde{x}_3,\tilde{y}_3)=f(k-2\tilde{x}_3)\tilde{y}_3$ with $k/2<\tilde{x}_3<k$ and $\tilde{y}_3>0$, then $f_{2x}(\tilde{x}_3,\tilde{y}_3)<0$ and $f_{2y}(\tilde{x}_3,\tilde{y}_3)<0$. By using the above calculations we obtain $f_{1x}f_{2y}<0$ and $f_{2x}f_{1y}>0$ then, the determinant of the Jacobian matrix of the system (\ref{E2}) at $P_3$

\[detJ[f_1,f_2]_{(\tilde{x}_3,\tilde{y}_3)}=(f_{1x}f_{2y}-f_{2x}f_{1y}),\]
is negative. Therefore $P_3$ is a saddle point of (\ref{E2}) for all the parameter values. 

For the proof of the first part of the Proposition we follow a similar sign analysis as we did previously, by considering that $0<\tilde{x}_2<k/2$, $\tilde{y}_2>0$ and that at $\tilde{x}_2$ the inequality $y_h'(\tilde{x}_2)<y_v'(\tilde{x}_2)$ holds where both derivatives are positive. Thus, given that $f_{1y}(\tilde{x}_2,\tilde{y}_2)=-c\tilde{x}_2(k-\tilde{x}_2)$ and $f_{2x}(\tilde{x}_2,\tilde{y}_2)=f(k-2\tilde{x}_2)\tilde{y}_2$ the inequalities

\[f_{1y}<0,\;\;f_{1x}>0\;\;\mbox{and}\;\;f_{2x}>0,\;f_{2y}<0,\]
follow, from which we get $f_{1x}f_{2y}<0$ and $f_{2x}f_{1y}<0$. By using these inequalities and the condition $y_h'(\tilde{x}_2)<y_v'(\tilde{x}_2)$ one obtains

\[detJ[f_1,f_2]_{(\tilde{x}_2,\tilde{y}_2)}=(f_{1x}f_{2y}-f_{2x}f_{1y})>0,\]
with this we complete the proof. $\;\;\;\Diamond $

\vspace{4mm} 
\noindent
{\bf Remark 1.} The trace of the Jacobian matrix (\ref{E9}) at any point 
$(x,y)$ is the quadratic

\begin{equation}
\label{E11}
trJ[f_1,f_2]_{(x,y)}= -fx^2+2cxy+(fk-2b)x-(ck+2e)y+b-e.
\end{equation}

Given that its discriminant\footnote{For the calculation of the 
discriminant we consider the general form of the quadratic 

\[Ax^2+2Bxy+Cy^2+Dx+Ey+F=0,\] 
hence $A=-f$, $B=c$ and $C=0$.} 
$\Delta=AC-B^2=-c^2<0$, (\ref{E11}) is a quadratic equation of hyperbolic 
type. In order the see more details of such quadratic, we calculate its 
gradient. This is the zero vector at the point

\[(\hat{x},\hat{y})=\left(\frac{ck+2e}{2c},\frac{fe+bc}{c^2}\right)\in{\mathcal R}.
\]
Given that the partial derivatives\footnote{For notational convenience we 
simply write $tr J$ instead of $trJ[f_1,f_2]_{(x,y)}$.}

\[\frac{\partial^2 trJ}{\partial x^2},\;\;\;\;\frac{\partial^2 
trJ}{\partial y\partial x}\;\;\;\mbox{and}\;\;\;\frac{\partial^2 
trJ}{\partial y^2}\]
evaluated at any point $(x,y)$ ---in particular $(\hat{x},\hat{y})$--- 
are: $-2f<0$, $2c>0$ and $0$ respectively, then we have

\[\left[\frac{\partial^2 trJ}{\partial x^2}\right]\left[\frac{\partial^2 trJ}{\partial y^2}\right]-\left\{\frac{\partial^2 trJ}{\partial y\partial x}\right\}^2=-4c^2<0.\]
Therefore $(\hat{x},\hat{y})$ is a saddle point of the surface 
(\ref{E10}). The value of $trJ[f_1,f_2]$ at $(\hat{x},\hat{y})$ is

If, for the parameter values we prove 
$trJ[f_1,f_2]_{(\tilde{x}_2,\tilde{y}_2)}<0$ in addition to what we already 
have had proved, our conclusion follows.

\newpage

%
%
%
%
%
%
%
%
%
%
%
%
%

\end{document}